\documentclass[aps,prb,amsmath,amssymb,showpacs,twocolumn,groupedaddress]{revtex4}

\usepackage[dvips]{graphicx}
\usepackage{bm}
\def\figwid{8cm}

\begin{document}

\preprint{Version 1.1}

\title{Dynamics of Spontaneous Magnetization Reversal in Exchange Biased Heterostructures}

\author{Zhi-Pan Li}
\email[E-mail: ]{zl65@cornell.edu}
\altaffiliation{Present address: Center of Nanoscale Systems, Cornell University, Ithaca, NY 14853.}
\author{Casey W. Miller}
\author{Igor V. Roshchin}
\author{Ivan\,K. Schuller}
\affiliation{Physics Department, University of California, San
Diego, La Jolla, CA, 92093-0319, USA}

\date{\today}

\begin{abstract}
The dependence of thermally induced spontaneous magnetization
reversal on time-dependent cooling protocols was studied.  Slower
cooling and longer waiting close to the N\`{e}el temperature of
the antiferromagnet ($T_N$) enhances the magnetization reversal.
Cycling the temperature around $T_N$ leads to a thermal training
effect under which the reversal magnitude increases with each
cycle. These results suggest that spontaneous magnetization
reversal is energetically favored, contrary to our present
understanding of positive exchange bias.
\end{abstract}


\maketitle

Exchange bias (EB) arises when a ferromagnet/antiferromagnet
(FM/AF) heterostructure is cooled in a magnetic field ($H_{FC}$)
below the N\'eel temperature $T_N$ of the AF \cite{Meiklejohn67,
Nogues2, Kiwi81}. EB is manifested as a shift of the hysteresis
loop along the field axis by an amount $H_{EB}$, dubbed the
exchange bias field. This phenomenon has been intensely studied in
the past ten years due to its significance in providing a magnetic
reference in spin valve devices\cite{Kools111}. More
fundamentally, EB is also of physical importance for understanding
competing interactions in coupled magnetic materials. A rich
variety of physical phenomena are associated with EB, including
thermal stability \cite{Skumryev110}, positive EB \cite{Nogues77},
training effect \cite{Paccard112, Hoffmann109}.

We recently demonstrated that a thin exchange biased FM layer can
fully reverse its magnetization to point against a constant
$H_{FC}$ during cooling \cite{Li108}. Similar behavior has been
observed in ferrimagnet Gd-Co \cite{CoPd1} and Co/Gd multilayer
systems \cite{CoPd2}, which results from two antiferromagnetically
coupled spin species competing to align with the field. In the
present case, a prerequisite for spontaneous magnetization
reversal is positive EB, where the hysteresis loop is shifted in
the direction of $H_{FC}$ \cite{Nogues77}. Positive EB arises when
$H_{FC}$ is large enough to overcome the antiferromagnetic
interfacial coupling, thus aligning uncompensated AF moments along
the field. In other words, positive EB requires
$|E_{int}|<|E_{AF-Zeeman}|$, where $E_{int}$ and $E_{AF-Zeeman}$
are the interfacial coupling energy and the Zeeman energy of
uncompensated AF moments, respectively. Negative EB arises when
$H_{FC}$ is small enough that the interfacial coupling forces AF
moments to align antiparallel with the field. When positively
exchange biased, a FM spontaneously reverses its magnetization
when $|E_{int}|>|E_{FM-Zeeman}|$ \cite{Li108}. This implies
$|E_{FM-Zeeman}|<|E_{AF-Zeeman}|$, or equivalently
$|m_{FM}|<|m_{AF}|$, where $m_{FM}$ and $m_{AF}$ are FM and
uncompensated AF magnetic moments, respectively. However, our
experiment indicated $|m_{FM}|>>|m_{AF}|$ because there was no
evidence of a significant vertical shift of the low-temperature
hysteresis loop (a signature of pinned AF moments), or a reduction
of the saturation magnetization (a signature of unpinned AF
moments) \cite{Nogues75}. Therefore, spontaneous magnetization
reversal seems energetically unfavorable within the existing
framework of exchange bias. A lower energy state would have
positive FM and negative AF moments, corresponding to negative EB,
rather than observed positive EB. This contradiction implies that
either spontaneous reversal is a novel metastable state, or our
present understanding of positive EB is incomplete.

This work reports slow dynamics and thermal training of the
spontaneous reversal effect. We show that a slow cooling rate
enhances the magnetization reversal magnitude, and that reversal
is strongly related to dynamic processes around $T_N$. Relaxation
of the system at $T_N$ over a long period of time causes increased
reversal at low temperatures. Successive thermal cycling about
$T_N$ allows the system reach a global equilibrium state. These
results show that spontaneous reversal is energetically favored
rather than a metastable state as predicted by the existing
positive exchange bias model. Possible directions for a modified
theory of positive exchange bias are suggested.

\begin{figure}
\centering
\includegraphics[width=4.2cm]{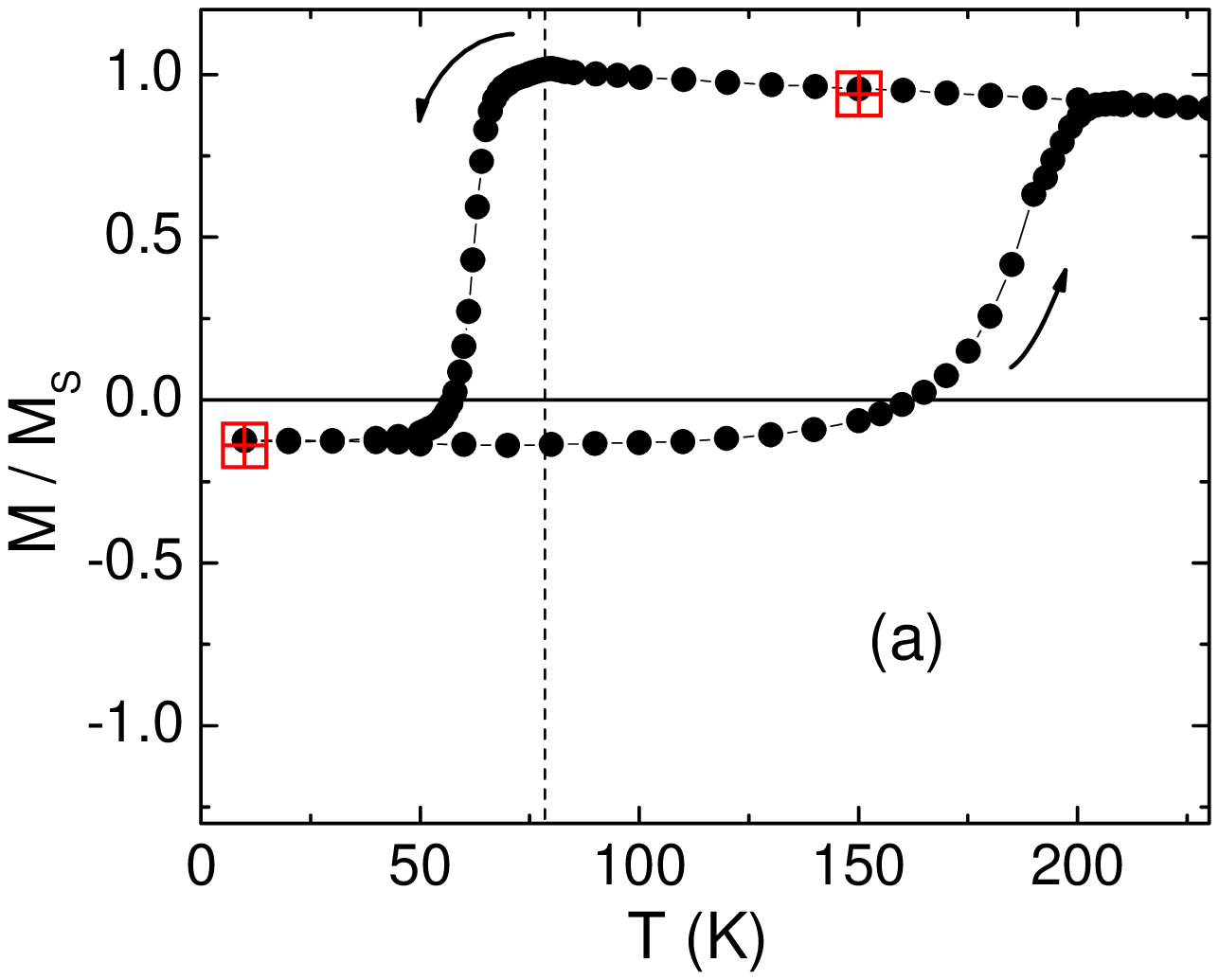}
\includegraphics[width=4.2cm]{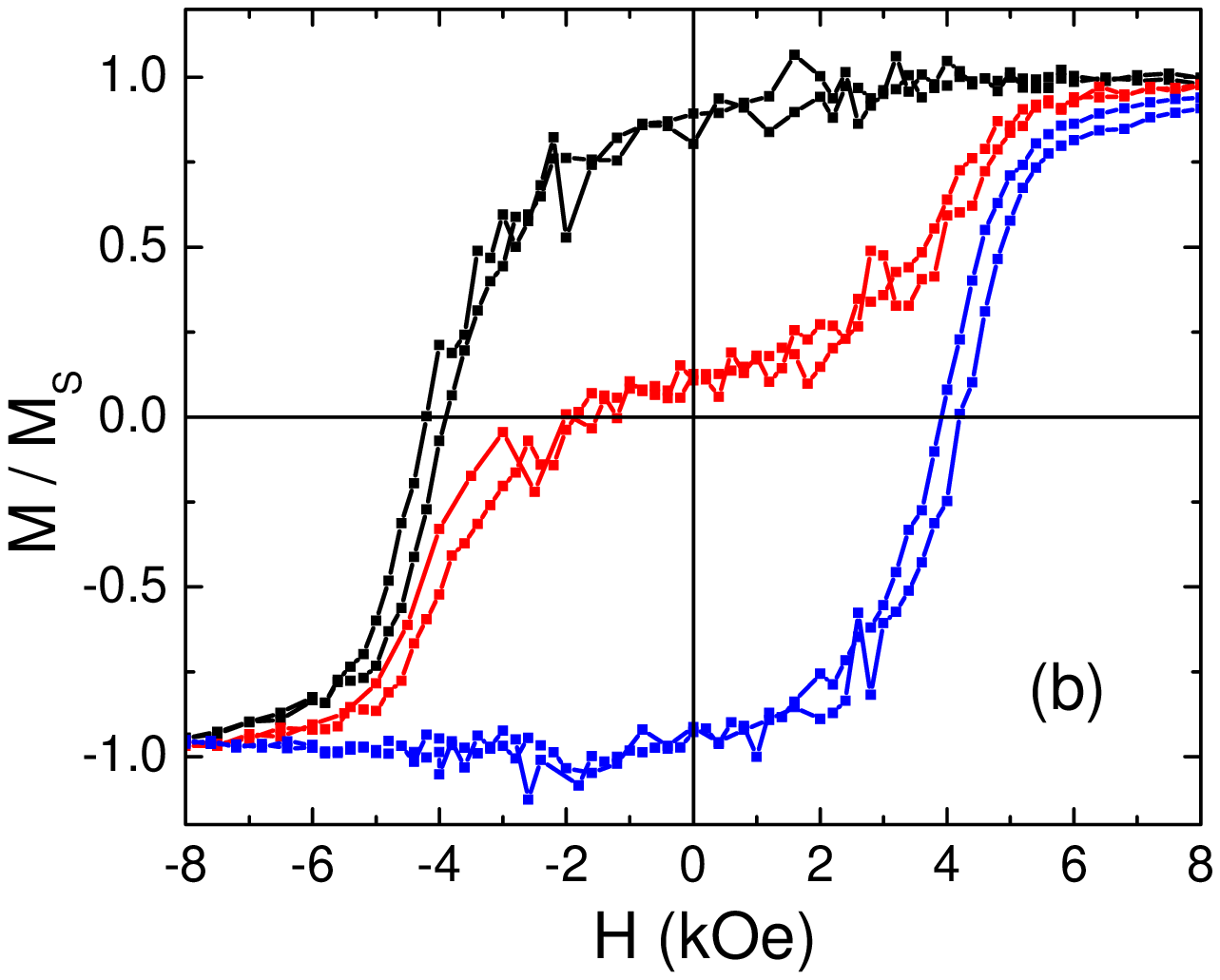}
\caption{(Color online)(a) $M\,vs\,T$ of Ni/FeF$_2$ normalized by
$M_S$ with $H_{FC}=0.1$\,kOe. The dashed line marks $T_N=78$\,K.
The red crossed squares mark the initial and final normalized
magnetization of the same cooling procedure from 150\,K to 10\,K
but without measurement at intermediate temperatures. (b) Magnetic
hysteresis at $T=10$\,K after cooling with $H_{FC}$ of -0.1\,kOe
(black), 0.1\,kOe (red) and 1\,kOe (blue) at
10\,K/min.}\label{hyst}
\end{figure}

The same Ni(3nm)/FeF$_2$(30nm) sample on a MgF$_2$ substrate was
studied as in our previous work \cite{Li108}. FeF$_2$ is an AF
with $T_N=78$\,K, and grows epitaxially untwinned in the (110)
direction on MgF$_2$ (110) substrates \cite{Petracic105}. The FM
exhibits uniaxial anisotropy with the easy axis parallel to
FeF$_2$ [001] (the spin axis of the AF). The magnetic field was
always applied along the easy axis of the FM. Prior to cooling,
the FM was saturated with a 5\,kOe field, well above the 150\,K
coercive field of $H_C=0.35$\,kOe, then reduced to $H_{FC}$. Fig.
\ref{hyst}(a) shows the thermally induced magnetization reversal
in $H_{FC}=0.1$\,kOe. Hysteresis measurements at $T=10$\,K find
negative EB for $-0.25\textrm{\,kOe} \le H_{FC} \le -0.1$\,kOe,
positive EB at $H_{FC}>0.5$\,kOe, and coexistence for -0.1
\textrm{\,kOe} $< H_{FC} <$ 0.5 kOe (Figure \ref{hyst}(b)).
$|H_{EB}|=(3.9\pm0.1)$\,kOe for all cooling fields. Coexistence of
positive and negative EB at an intermediate $H_{FC}$ has been
interpreted as the AF breaking into ``domains" with uncompensated
moments of either sign \cite{Petracic105}. When the lateral size
of these ``domains" is much larger than the FM domain wall width,
they independently induce either positive or negative EB in the
FM, causing the experimentally observed double hysteresis loop.
Since only positive EB is essential for spontaneous reversal,
partial reversal was observed for $H_{FC}$ associated coexistence
(Fig. \ref{hyst} (a)).

\label{sec:spd}\begin{figure}
\centering
\includegraphics[width=\figwid]{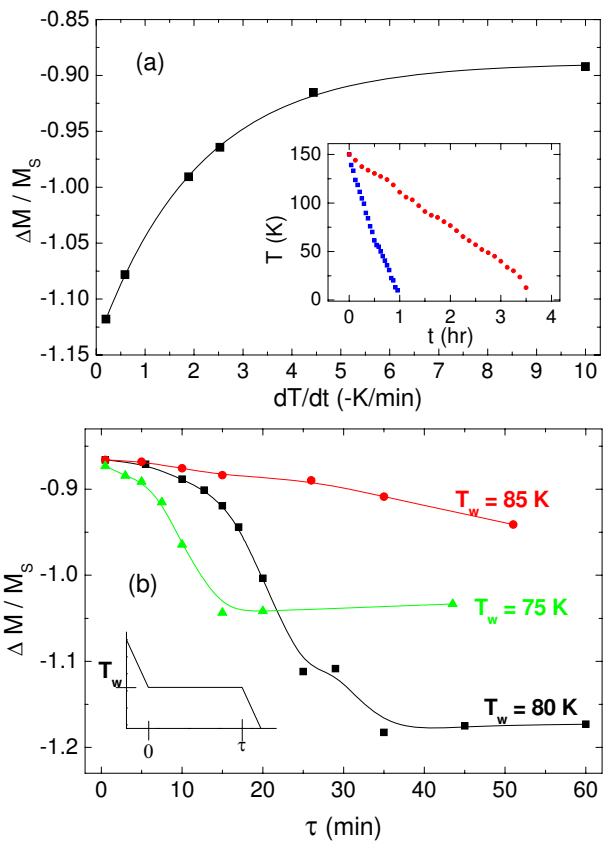}
\caption{(Color online) (a) $\Delta M/M_S$ as function of uniform cooling speed
$dT/dt$ for $H_{FC}=0.1$\,kOe. The line is a fit to an exponential
function. (Inset) Temperature $vs$ time for uniform cooling speeds
of 2.5 (blue) and 0.6\,K/min (red). (b) $\Delta M/M_S$ as a function of wait time
$\tau$ at temperatures $T_w=75$(green), 80(black), and 85(red)\,K
for $H_{FC}=0.1$\,kOe. (Inset) Schematic of this cooling protocol.
Lines are a guide to the eye.}\label{waittime}\label{coolingspd}
\end{figure}

Two different cooling protocols were used to investigate the time
dependence of the reversal magnitude $\Delta M=M(T=10 \text{
K})-M(T=150 \text{ K})$. We only consider $H_{FC}=0.1$\,kOe, for
which the magnetization reverses by about 50\% upon cooling. The
first protocol cooled the sample from $T=150$\,K to 10\,K with two
decades (0.1-10 K/min) of uniform cooling speeds (Fig.
\ref{coolingspd}(a) inset). The second protocol cooled the sample
at 10\,K/min from 150\,K to an intermediate temperature $T_w$,
where the temperature was held constant for a time $\tau$, then
cooled to 10\,K at 10\,K/min (Fig. \ref{waittime}(b) inset). For
both protocols, $M(T=10 \textrm{ K})$ was measured after the
sample temperature stabilized at 10\,K.

Fig. \ref{coolingspd} (a) shows that slower cooling leads to a
larger $|\Delta M|$.  With the present definition, $\Delta M=-2$
implies complete magnetization reversal.  With the largest cooling
speed of 10\,K/min, the FM reverses by $\Delta M=-0.9M_S$. When
cooled at 0.1\,K/min, $|\Delta M|$ increases by $0.2M_S$.
Moreover, the dependence of $\Delta M$ on the cooling speed is
well fit by an exponential function $\Delta M=\Delta M_0 + A
\exp(\alpha dT/dt)$. This fit implies $\Delta M$ of -0.88 and
$-1.15 M_S$ in the limits of infinite and zero cooling speed,
respectively.

The second cooling protocol demonstrates that $|\Delta M|$ is
sensitive to the time spent with $T\sim T_N$. The dependence of
$\Delta M$ on the wait temperature $T_w$ shows the largest
reversal for $T_w=80$\,K, closest to $T_N$ (Fig. \ref{waittime}).
As $\tau$ increases beyond 35 min, $|\Delta M|$ increases from
$0.86M_S$ to $1.18M_S$. For $T_w=85$\,K, $|\Delta M|$ only changes
by $0.07M_S$ after waiting for 50 minutes. For $T_w=75$\,K,
$|\Delta M|$ saturates after $\sim15$ minutes at $1.05M_S$. This
$T_w\textrm{-dependent}$ behavior was not observed when waiting at
the reversal temperature ($T=63$\,K). The results from these two
cooling protocols show that spontaneous magnetization reversal
exhibits slow dynamics with a relatively long time scale. The fact
that the dynamics are most pronounced around $T_N$ hints that this
effect depends on the establishment of AF domain states.

\begin{figure}
\includegraphics[width=\figwid]{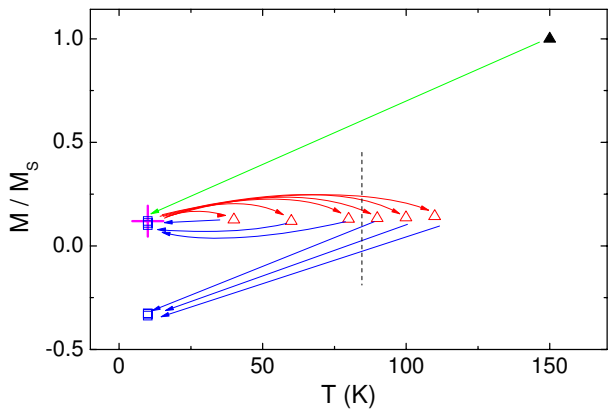}
\caption{(Color online) Normalized magnetization $M/M_S$ was
measured after each step of the three-step thermal cycle: (1) cool
from $T=150$\,K to 10\,K (magenta cross); (2)
warm to $T_x$ (40-110\,K) (red triangles); (3) cool back down to
10\,K (blue squares). $M/M_S$ was reversible for $T_x<80$ K, but
irreversible above 80\,K. The dashed line separates these two
regimes, and is close to $T_N$. The lines are schematics of the
measurement sequence.}\label{chk}
\end{figure}

Several tests were performed to ensure that the dynamics were not
experimental artifacts. First, measuring the magnetic moment via
SQUID involves moving the sample through the SQUID coils by 4\,cm,
thus subjecting the sample to magnetic field inhomogeneity. To
exclude this as an artifact, the cooling procedure used to obtain
the data of Fig. \ref{hyst} was repeated, but measuring only the
initial and final the magnetization values, rather than at several
intermediate temperatures.  The sample was thus only exposed to
field inhomogeneity at these extreme temperatures. The reversal
magnitude only differs by $3\times10^{-4}M_S$ between these two
methods, which is negligibly small. Temperature fluctuations
during cooling are another source of potential signal artifacts.
To investigate this, the sample was heated from 10\,K to a
temperature $T_x$, then cooled back down to 10\,K. When $T_x \le
80$\,K, $\Delta M$ varies by no more than $0.01M_S$, too small to
account for any $\Delta M$ variation found earlier (Fig.
\ref{chk}). When $T_x>80$\,K, a significant additional
magnetization reversal was observed (more below). These checks
demonstrate that the time-sensitivity of spontaneous reversal is
not an experimental artifact but rather is intrinsic to the
system, and clearly related to the AF phase transition. Larger
reversal for slower cooling rates and longer wait times around
$T_N$ suggests that spontaneous reversal is thermodynamically
favorable.

\begin{figure}
\includegraphics[width=4.2cm]{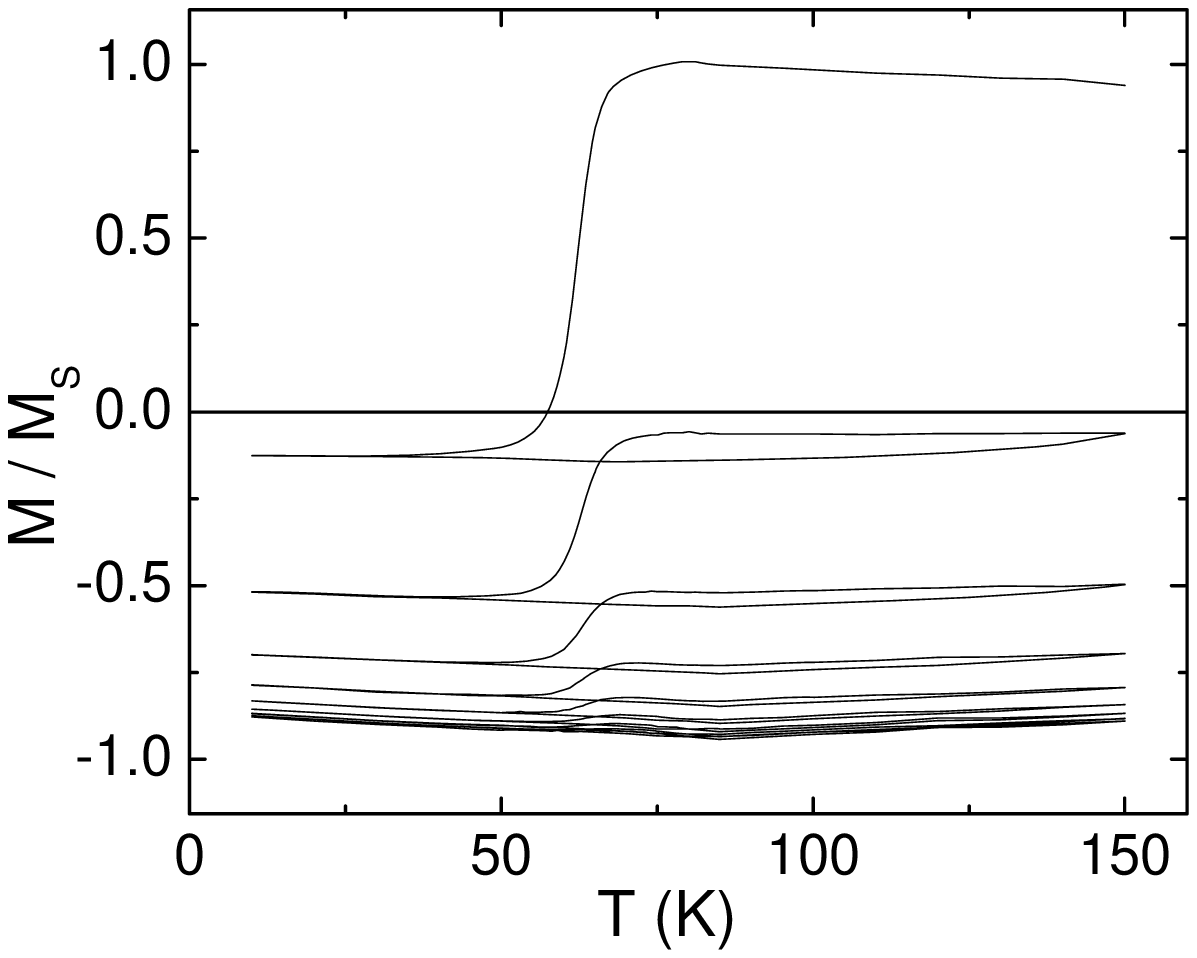}
\includegraphics[width=4.2cm]{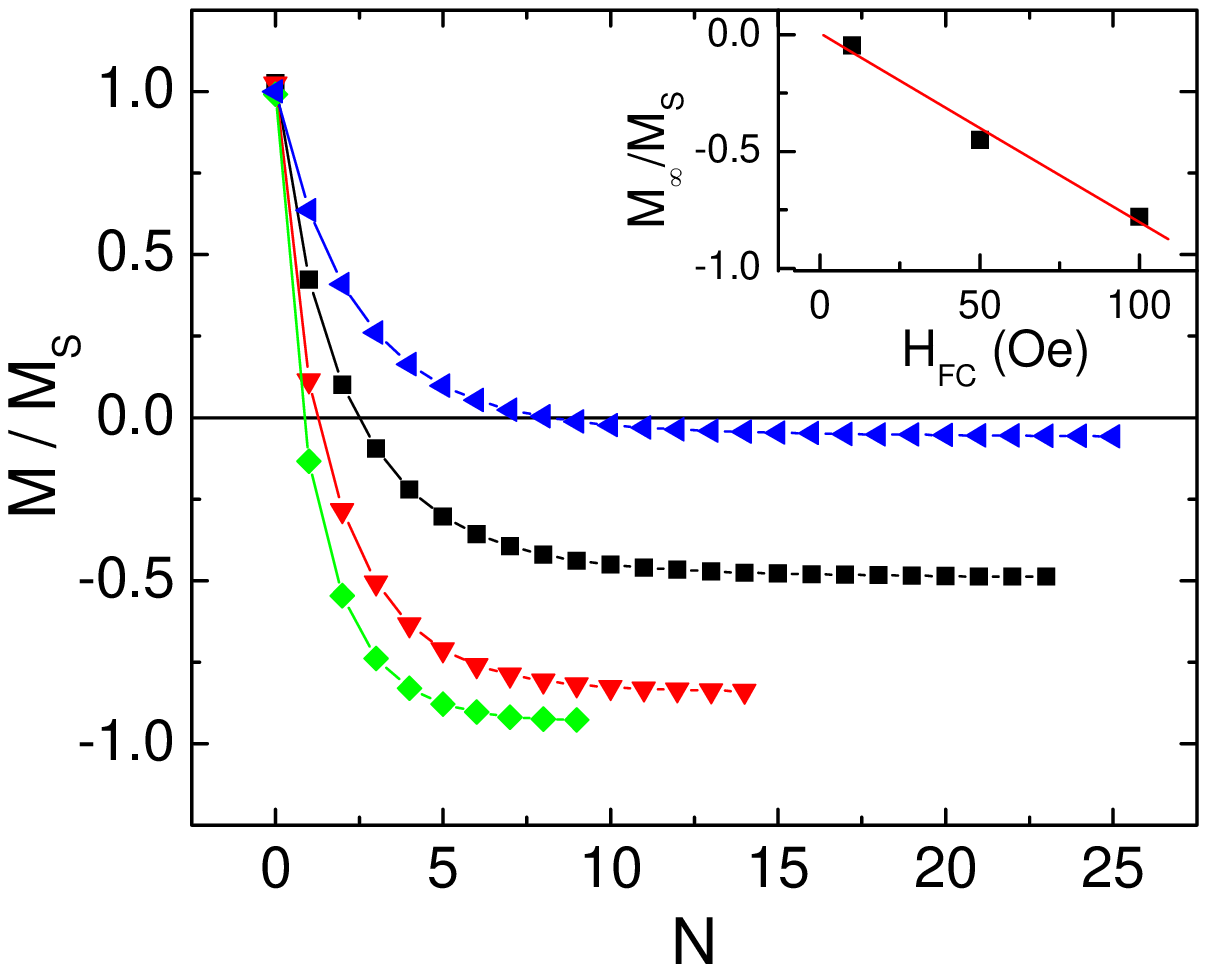}
\caption{(Color online) (a) Normalized magnetization $M/M_S$ was measured as the
temperature was cycled between 150\,K and 10\,K in a 0.1\,kOe
field. Nine cycles were conducted. (b) Normalized magnetization $M/M_S$ at $T=10$ K after
each thermal cycle between 150\,K and 10\,K as a function of the
number of cycles $N$. Different cooling parameters were used,
$H_{FC}=10$\,Oe(blue), 50\,Oe(black), 100\,Oe(red) with 10\,K/min
cooling/heating rate, and $H_{FC}=100$\,Oe with 0.1\,K/min
(green). The lines are a guide to the eye. (Inset) Asymptotic
magnetization $M_{\infty}/M_S$ at high cycling number $N$ obtained
by exponential fitting, as a function of $H_{FC}$. The red line is
a linear fit.}\label{decay} \label{cycle}
\end{figure}

The slow evolution of the system toward a larger reversal
implies the presence of large energy barriers. The additional
large reversal during thermal cycling above 80\,K (Fig. \ref{chk})
suggests that the system can overcome this energy barrier by
thermal activation. \emph{Thermal training} refers to successive
magnetization reversal when the system is cycled above and below
$T_N$. With the FM saturated, the sample was first cooled in
$H_{FC}=0.1$\,kOe from 150\,K to 10\,K at 0.1\,K/min, followed by
heating to 150\,K, just below the temperature for the FM to
reverse back along the field direction (Fig. \ref{hyst}). After that, the sample was
cycled between 150\,K and 10\,K. The magnetic field was held
constant at $H_{FC}=0.1$\,kOe throughout the thermal cycles. The
FM reverses with each additional cooling (with decreasing
incremental reversal magnitude) until the total magnetization
reversal reaches 1.8$M_S$, significantly larger that
the initial reversal for any cooling speed or wait time (Fig. \ref{cycle}(a)).

Fig. \ref{decay}(b) shows the dependence of $M(10\,\text{K})$
on the number of cycles $N$ for different $H_{FC}$ and cooling
speeds. For all cases, they follow an exponential dependence,
$M_N(10\,\text{K})=M_{\infty}+(M_S-M_{\infty})\exp(-N/\eta)$,
where $M_{\infty}$ is the convergent $M(10\,\text{K})$ when
$N\rightarrow\infty$, and $\eta$ is a characteristic cycle number
for each $H_{FC}$ and cooling speed. $M_N(10\,\text{K})$ for $N=0$
is defined as $M_S$. $M_{\infty}$ appears to be linearly dependent
on $H_{FC}$ for constant cooling speed (Fig. \ref{decay} inset).
Larger $H_{FC}$ results in smaller $\eta$, which means a faster
approach to $M_{\infty}$. This makes qualitative sense because a
larger magnetic field should facilitate reversal by lowering the
energy barrier, so that more AF moments are aligned in the field direction.


These experiments suggest that it is energetically favorable for
the FM to reverse against $H_{FC}$, albeit counterintuitive since
$|m_{FM}|>>|m_{AF}|$. This behavior cannot be explained simply by
the competition between the Zeeman energy and interfacial
coupling. A new mechanism for determining the sign of AF
uncompensated moments is necessary to explain the features we
observe experimentally. Consider that $(M_S-M(10\,\text{K}))/2M_S$
gives the percentage of sample that exhibits positive EB at 10\,K
for an intermediate $H_{FC}$. For $H_{FC}=0.1$\,kOe, the sample is
nearly 90\% positively exchange biased at 10\,K after 6 thermal
cycles at 0.1\,K/min. The interfacial coupling energy in this
sample is
$E_{int}=J_{FM/AF}\mathbf{S}_{FM}\cdot\mathbf{S}_{AF}=\mu_0H_{EB}M_{FM}t_{FM}=0.79\text{
erg/cm}^2$, close to that previously found in similar systems
\cite{Nogues76}. However, the onset $H_{FC}$ for positive EB in
this case is about two orders of magnitude smaller than previously
found \cite{Leighton72}. This very small $H_{FC}$ necessary for
positive EB challenges the present interpretation of positive EB.

One possible explanation for these observations considers pinned
uncompensated moments in the bulk of the AF. Neutron scattering
results show that parallel AF domain walls can form between the
bulk and interfacial AF moments when the interfacial moments are
more strongly coupled with the FM \cite{Roy104}. For small (large)
cooling fields, interfacial AF moments need to orient in the
negative (positive) direction to establish negative (positive) EB.
Here, antiferromagnetic interfacial coupling is assumed. However,
bulk AF moments far away from the interface are dominated by the
applied field and align parallel to it. This is independent from
the orientation of the interfacial AF moments. Therefore, a
parallel AF domain wall forms in case of negative EB, but not for
positive EB.

A second scenario considers a parallel domain wall in the FM.
Spontaneous rotation found previously supports this possibility
\cite{Li108}. When cooling down, such a domain wall occurs in
positively exchange biased thick FMs because the antiferromagnetic
interfacial coupling locks interfacial FM moments in the negative
direction while FM moments far from the interface only sense the
external field \cite{Li117, Morales163}. With the inclusion of the
parallel AF and/or FM domain wall energies, the sign of exchange
bias is no longer determined simply by the competition between
$|E_{int}|$ and $|E_{AF-Zeeman}|$, and the paradox
$|m_{FM}|<<|m_{AF}|$ can be avoided.
However, detailed calculations of the different energies involved
in negative and positive EB are necessary to explicitly develop
the pertinent relationships.

The observed slow dynamics may arise from the competition of these
energies around $T_N$, which also determine the sign of EB while
the AF order is established. The competition of these energies may
result in multiple local energy minima that are separated by
significant anisotropy barriers. These barriers grow larger
compared with $k_BT$ with decreasing temperature, and the time it
takes for the system to evolve into a lower energy state
exponentially increases. Thermal training allows the system to
seek out the global energy minimum because the FM domains are
approximately unchanged with temperature, but the AF order is
cyclicly perturbed. After cooling to $T_x<T_N$ for the first time,
the AF orders, and a portion of the originally saturated FM
spontaneously reverses because $|E_{int}|>|E_{FM-Zeeman}|$. Next,
the AF becomes disordered when warmed to $T_x>T_N$ with the FM
domains relatively unchanged. Subsequent cooling causes the
population of positive EB regions to increase because the FM
moments associated with the domain walls deviate from the field
direction. This results in a smaller coupling energy with the AF,
making it easier for the AF moments to align with the field. This
gives rise to a larger fraction of the sample that shows positive
EB, and thereby increases the magnetization reversal magnitude.
This process is successful because the FM domain wall width is an
order of magnitude larger than the AF domain wall width. The
details of the time-dependent reversal, what determines the system
ground state, and the various paths to reach this state are not
presently understood.

In summary, two different cooling protocols revealed that
spontaneous magnetization reversal in exchange biased
heterostructures is strongly time-dependent. Slower cooling speeds
and longer waiting times around $T_N$ lead to larger magnetization
reversal. Thermal training was discovered by cycling the sample
temperature about $T_N$, causing the FM to reverse successively
with each cycle.  This effect reflects the incremental conversion
of negative to positive EB regions. These results suggest that
spontaneous reversal is thermodynamically stable rather than
metastable, contradicting our present understanding of positive
EB. Additional energy terms that describe parallel domain walls in
the antiferromagnet and/or ferromagnet are necessary to explain
these results, and to refine positive exchange bias models.

\vspace{0.3in} This work was supported by US-DOE. The authors
thank T.\,Gredig for illuminating discussions.

\end{document}